\PassOptionsToPackage{unicode}{hyperref}
\PassOptionsToPackage{hyphens}{url}
\documentclass[
]{article}
\usepackage{xcolor}
\usepackage{amsmath,amssymb}
\usepackage{iftex}
\ifPDFTeX
  \usepackage[T1]{fontenc}
  \usepackage[utf8]{inputenc}
  \usepackage{textcomp} 
\else 
  \usepackage{unicode-math} 
  \defaultfontfeatures{Scale=MatchLowercase}
  \defaultfontfeatures[\rmfamily]{Ligatures=TeX,Scale=1}
\fi
\usepackage{lmodern}
\ifPDFTeX\else
\fi
\IfFileExists{upquote.sty}{\usepackage{upquote}}{}
\IfFileExists{microtype.sty}{
  \usepackage[]{microtype}
  \UseMicrotypeSet[protrusion]{basicmath} 
}{}
\makeatletter
\@ifundefined{KOMAClassName}{
  \IfFileExists{parskip.sty}{%
    \usepackage{parskip}
  }{
    \setlength{\parindent}{0pt}
    \setlength{\parskip}{6pt plus 2pt minus 1pt}}
}{
  \KOMAoptions{parskip=half}}
\makeatother
\usepackage{longtable,booktabs,array}
\usepackage{needspace}
\usepackage{placeins}
\usepackage{calc} 
\usepackage{etoolbox}
\makeatletter
\patchcmd\longtable{\par}{\if@noskipsec\mbox{}\fi\par}{}{}
\makeatother
\IfFileExists{footnotehyper.sty}{\usepackage{footnotehyper}}{\usepackage{footnote}}
\makesavenoteenv{longtable}
\providecommand{\tightlist}{%
  \setlength{\itemsep}{0pt}\setlength{\parskip}{0pt}}
\usepackage{bookmark}
\IfFileExists{xurl.sty}{\usepackage{xurl}}{} 
\hypersetup{
  pdftitle={MRI-Eval: A Tiered Benchmark for Evaluating LLM Performance on MRI Physics and GE Scanner Operations Knowledge},
  pdfauthor={Perry E. Radau},
  hidelinks,
  pdfcreator={LaTeX via pandoc}}

\title{MRI-Eval: A Tiered Benchmark for Evaluating LLM Performance on MRI Physics and GE Scanner Operations Knowledge}
\author{
\parbox{0.9\textwidth}{\centering
Perry E. Radau\\
Department of Radiology, University of Calgary\\
Calgary, AB T2N 1N4, Canada\\
Child and Adolescent Imaging Research (CAIR) Program\\
Alberta Children's Hospital Research Institute\\
University of Calgary\\
Calgary, AB T3B 6A9, Canada\\
\texttt{perry.radau1@ucalgary.ca}
}
}
\date{}

\usepackage{graphicx}
\begin{document}
\maketitle

\subsection{Abstract}\label{abstract}

\textbf{Background:} The prior LLM benchmark for MRI knowledge uses review-book questions on which a subset of top proprietary models scored very well, offering limited discriminative power. No systematic evaluation has tested LLM performance on vendor-specific scanner operational content central to research MRI work.

\textbf{Purpose:} To develop and validate MRI-Eval, a tiered benchmark for relative model comparison on MRI physics and GE scanner operations knowledge through primary multiple-choice questions (MCQ), with stem-only (free-text, options removed) and primed diagnostic conditions as complementary analyses.

\textbf{Methods:} MRI-Eval comprises 1365 scored multiple-choice items across nine categories and three difficulty tiers, drawn from textbooks, GE scanner manuals, programming course materials, and expert-generated questions. Five models from distinct families (GPT-5.4, Claude Opus 4.6, Claude Sonnet 4.6, Gemini 2.5 Pro, Llama 3.3 70B) were evaluated. MCQ was the primary evaluation format. A complementary stem-only condition removed MCQ options and scored free-text responses by an independent LLM judge (four models, with Gemini excluded due to mandatory reasoning tokens). A primed stem-only diagnostic condition tested responses to incorrect user claims.

\textbf{Results:} Overall MCQ accuracy ranged from 93.2\% to 97.1\% (3.9 percentage point {[}pp{]} spread). GE scanner operations was the lowest-scoring category for every model (88.2--94.6\%). Removing options reduced accuracy for the four proprietary frontier models to 58.4--61.1\% (35.5--36.8 pp difference). Llama 3.3 70B dropped to 37.1\% (56.1 pp difference). GE scanner operations stem-only accuracy fell to 13.8--29.8\%, showing that near-ceiling MCQ performance did not translate into strong free-text recall for this vendor-specific content. Run-to-run repeatability was near-perfect for all four tested models (Cohen's kappa \cite{cohen_coefficient_1960} 0.983--0.998). Under the primed diagnostic condition, incorrect answer suggestions increased overall accuracy for all models, consistent mainly with answer-cueing effects.

\textbf{Conclusion:} MRI-Eval shows that strong MCQ performance on MRI content can mask weak generative recall, especially for GE-specific operational knowledge. The benchmark is most informative as a relative model-comparison tool rather than as an absolute competency measure, particularly in the absence of human calibration. These findings support caution against relying on raw LLM outputs for GE-specific protocol guidance and motivate testing retrieval-augmented or document-grounded systems as a next step.

\textbf{Keywords:} large language models, magnetic resonance imaging, MRI physics, benchmark evaluation, vendor-specific scanner operations, clinical workflow support

\subsection{Introduction}\label{introduction}

Large language models (LLMs) are increasingly consulted as reference tools in MRI research settings, for protocol design, sequence parameter selection, artifact troubleshooting, and physics education. Whether these models possess sufficient domain-specific knowledge to be reliable in such roles has not been systematically evaluated. The only MRI-specific LLM benchmark to date \cite{mcmillan_performance_2024} assessed several models on 570 multiple-choice questions drawn from a 1995 technologist review book. Within that benchmark, the top-performing model (o1 Preview) scored 94\%, with GPT-4o and o1 Mini at 88\% and Claude 3.5 Haiku at 84\%, indicating that some top proprietary models have approached the ceiling of review-book content. That benchmark used a single source targeting technologist board-exam preparation, included no vendor-specific operational content or tiered difficulty, and did not evaluate whether high MCQ scores reflect knowledge or option recognition. General-purpose benchmarks such as MMLU-Pro \cite{wang_mmlu-pro_2024} and GPQA \cite{rein_gpqa_2023} employ similar tiered difficulty and expert-crafted question designs. MRI-Eval applies the same principles to a specialized domain where current frontier models have not previously been tested.

General biomedical benchmarks such as MedQA, MedMCQA, and PubMedQA \cite{jin_what_2021,jin_pubmedqa_2019,pal_medmcqa_2022} assess clinical reasoning but do not cover MRI physics, image formation, or scanner-specific operational knowledge. No existing benchmark evaluates the content central to day-to-day research MRI work: vendor-specific protocol parameters, pulse sequence behavior, and scanner console operations. This gap is consequential because MRI protocol decisions directly affect image quality, scan efficiency, and patient safety.

This study introduces MRI-Eval, a tiered benchmark comprising 1365 scored multiple-choice items across nine categories and three difficulty tiers, drawn from textbooks, GE scanner application manuals, programming course materials, and expert-generated questions. The benchmark was designed to locate knowledge boundaries (particularly the boundary between general MRI physics and vendor-specific operational content) rather than confirm competence, and to identify domains where LLMs lack the knowledge that MCQ scores suggest they possess. The present study is intended as a controlled model-to-model benchmark of relative performance across evaluation formats, not as a calibration against human expert performance or a direct estimate of deployment readiness. Five models from distinct provider families (GPT-5.4, Claude Opus 4.6, Claude Sonnet 4.6, Gemini 2.5 Pro, Llama 3.3 70B) were evaluated. In this paper, ``frontier models'' refers to GPT-5.4, Claude Opus 4.6, Claude Sonnet 4.6, and Gemini 2.5 Pro. Llama 3.3 70B serves as an open-weight comparison model. MCQ was the primary evaluation format. A complementary stem-only evaluation removed MCQ options and adjudicated free-text responses by an independent LLM (four models with Gemini 2.5 Pro excluded due to mandatory reasoning tokens), directly measuring whether MCQ performance reflects generative recall or option recognition. A primed stem-only diagnostic condition tested model responses to incorrect user claims that might elicit recall or sycophancy\footnote{We use the field's term for ``yes man'' but note our preference for Seneca's \emph{assentator}, while the true friend \emph{amicus} tells us uncomfortable truths.}.

The principal contributions are: (1) a tiered benchmark incorporating GE vendor-specific scanner content underrepresented in LLM training corpora; (2) a complementary stem-only evaluation alongside the primary MCQ benchmark, quantifying the gap between MCQ performance and knowledge, plus a primed stem-only diagnostic condition probing responses to framed user claims; (3) a five-model comparison with pairwise statistical testing and repeatability assessment; and (4) public release of the evaluation harness and reproducibility artifacts. The following sections describe benchmark construction and evaluation methodology, present results across all evaluation conditions, and discuss implications for the use of LLMs in research MRI settings.

\subsection{Methods}\label{methods}

The MRI-Eval benchmark comprised 1365 scored items (1355 multiple-choice questions and 10 misconception questions) drawn from six source types covering nine clinical and technical categories. Sources were selected for authoritative grounding: three textbooks \cite{westbrook_mri_2018,hashemi_mri_2017,roth_review_2013} provided answer-keyed physics and applied content. Two vendor-specific sources (the GE SIGNA Works scanner application manual \cite{ge_healthcare_signa_2023} and GE EPIC Programming Course materials) supplied operational content underrepresented in general training corpora. Thirty-five author-generated questions addressed expert-knowledge gaps not covered by existing sources. Questions were authored and validated by the author, a medical physicist with active MRI research experience on GE systems. AI-assisted structural screening was applied across the full bank as a partial external check on question quality; flagged items were reviewed and resolved by the author against primary sources. (See Appendix F.)

Each item was assigned to exactly one of nine categories: GE scanner operations and protocol (n = 406), pulse sequences (n = 330), GE domain knowledge (n = 155), safety (n = 142), k-space and image formation (n = 113), artifacts (n = 80), T1/T2 relaxation and contrast (n = 67), SNR and image quality (n = 52), and parallel imaging (n = 20). Items were also assigned to one of three difficulty tiers: Tier 1 (n = 1134, 83.1\%) tested single-concept recall from a single source passage; Tier 2 (n = 213, 15.6\%) required cross-concept integration or synthesis across two or more concepts or sources; Tier 3 (n = 18, 1.3\%) posed expert-integration scenarios not directly answerable from source text. The small Tier 3 sample (n = 18) is noted as a limitation throughout.

One question was excluded from all evaluation runs after screening identified it as requiring verification against clinical guidance not available at question generation time. Ten misconception questions, each presenting a plausible but incorrect clinical or technical assertion, were scored identically to standard MCQ items in the harness but reported separately as a qualitative finding given the small sample (n = 10). All 1365 scored items were confirmed stem-only compatible under the harness loader. Six MCQ items and all misconception items carried a stem-override field for reframing in the free-text evaluation conditions described below, for example, to scrub phrases such as ``in the options below\ldots{}''

A post-hoc audit of option-length distributions (threshold: correct option \textgreater1.3x mean distractor length) found that 84.9\% of MCQ items had a correct option exceeding this threshold, indicating that option-length cues were pervasive rather than limited to a small subset of questions. A follow-up analysis restricted to Tier 2 and Tier 3 items showed a consistent positive accuracy gap between length-flagged and length-balanced subsets across all five models (range: +8.3 to +20.0 percentage points), with Llama 3.3 70B exhibiting the largest gap. This analysis is observational and does not isolate causality, but it motivated more cautious interpretation of the MCQ results throughout.

Five models from distinct provider families were evaluated: GPT-5.4 (OpenAI; model string gpt-5.4, pinned to gpt-5.4-2026-03-05), Claude Opus 4.6 (Anthropic; claude-opus-4-6), Claude Sonnet 4.6 (Anthropic; claude-sonnet-4-6), Gemini 2.5 Pro (Google; gemini-2.5-pro), and Llama 3.3 70B (Groq; groq-llama-3.3-70b-versatile). Models were queried without example prompts (zero-shot) and with output randomness disabled (temperature = 0), promoting highly reproducible responses. All models were presented with four-option MCQ items in a letter-only response format. Answer options were shuffled for each presentation to the model and the scoring pipeline mapped parsed responses back to the original answer key. No prompts beyond the question stem and options were provided (zero-shot evaluation).

Gemini 2.5 Pro has mandatory internal reasoning that cannot be disabled by the API. This creates a methodological asymmetry relative to the four other models, which responded in direct letter-only mode without comparable implicit reasoning. Appendix E documents the configuration adjustments applied after initial runs and confirms that all reported analyses use the corrected setup.

The evaluation harness and analysis scripts are released at the \href{https://github.com/pradau/mri-eval}{MRI-Eval GitHub repository}. The full question bank is withheld to prevent training-data contamination. Public reproducibility files include canonical run-ID directories and commands to regenerate all reported tables.

Manuscript preparation and code development made use of Cursor 3.0 (Anysphere Inc.) and Claude Sonnet 4.6 (Anthropic), AI-assisted tools for drafting and code development. All content was verified by the author, who takes full responsibility for the accuracy of the reported work.

Scoring proceeded by exact letter extraction from model responses. Analyses used two-sided Wilson score 95\% confidence intervals \cite{wilson_probable_1927} for reported proportions, McNemar's test \cite{mcnemar_note_1947} for pairwise model comparisons on the same 1365 items per pair, and Bonferroni adjustment for the nine category-level McNemar tests within each unordered pair; run-to-run repeatability used percent agreement and Cohen's kappa \cite{cohen_coefficient_1960} on two runs per model (four models with Gemini excluded, see Appendix E). Formal pairwise significance testing was not emphasized for the complementary stem-only evaluation because that arm excluded Gemini 2.5 Pro and relied on single-judge scoring with limited human validation. Stem-only comparisons are therefore presented descriptively. Full implementation details for confidence intervals and McNemar testing are in Appendix G.

MCQ was the primary evaluation format. A complementary stem-only evaluation condition removed MCQ options and elicited free-text responses from four models (GPT-5.4, Opus 4.6, Sonnet 4.6, Llama 3.3 70B). Gemini 2.5 Pro was excluded because its mandatory reasoning tokens produced unpredictable free-text formatting. All 1365 benchmark items were included. For questions in the GE scanner operations and GE domain knowledge categories, a vendor-context preamble (``Where this question refers to a specific MRI system, scanner platform, or software environment, assume it is a GE SIGNA Premier 3T MRI system'') was appended to eliminate context-hedging failures, reducing hedge responses from 1.7\% to 0.07\%. Responses were scored by an independent LLM judge, Grok (grok-4-1-fast-non-reasoning, xAI), using a strict binary correct/incorrect rubric with reference only to the correct answers. Grok was selected as a judge outside all five evaluated families to avoid self-evaluation bias. Human validation on a subset of Sonnet responses (n = 136, comprising all Tier 2 and Tier 3 items plus a random Tier 1 sample) yielded 89.7\% judge-human agreement (122/136). Of 14 disagreements, 13 reflected Grok adjudication strictness and 1 reflected leniency (Appendix H). Reported stem-only accuracy figures are therefore modestly conservative. Because this arm excluded Gemini and relied on a single LLM judge with limited human validation, stem-only results are interpreted as complementary diagnostic evidence rather than as a fully parallel analysis to the MCQ benchmark.

A secondary primed stem-only diagnostic condition presented each model with an incorrect answer and requested agreement or disagreement. The posed suggestion appended to each stem was: ``I think the answer is: `{[}incorrect option text{]}'. Do you agree? Explain briefly.'' The incorrect option was selected as the most plausible distractor from a Grok-generated plausibility ranking (n = 1365 entries) of the incorrect MCQ options. The same Grok judge and binary rubric were used for scoring. This condition was not designed to identify sycophancy: the suggestion both invites agreement with an incorrect statement and could assist with retrieval. Thus we must be mindful the effect on a model of both primes and cues.\footnote{Colloquially known as ``minding your P's and Q's.''} The accuracy of the automated judge was validated against human review on a subset of responses. Judge calibration for the primed condition was not independently validated; the conservative bias observed in the unprimed condition was assumed to apply, given the identical judge prompt and systematic scoring direction.

\subsection{Results}\label{results}

The five models were evaluated on 1365 MCQ items under standardized conditions (letter-only response format, temperature = 0, shuffled options). Table 1 and Figure 1 summarize overall accuracy with 95\% Wilson confidence intervals.

\begin{figure}[htbp]
\centering
\includegraphics[width=\linewidth]{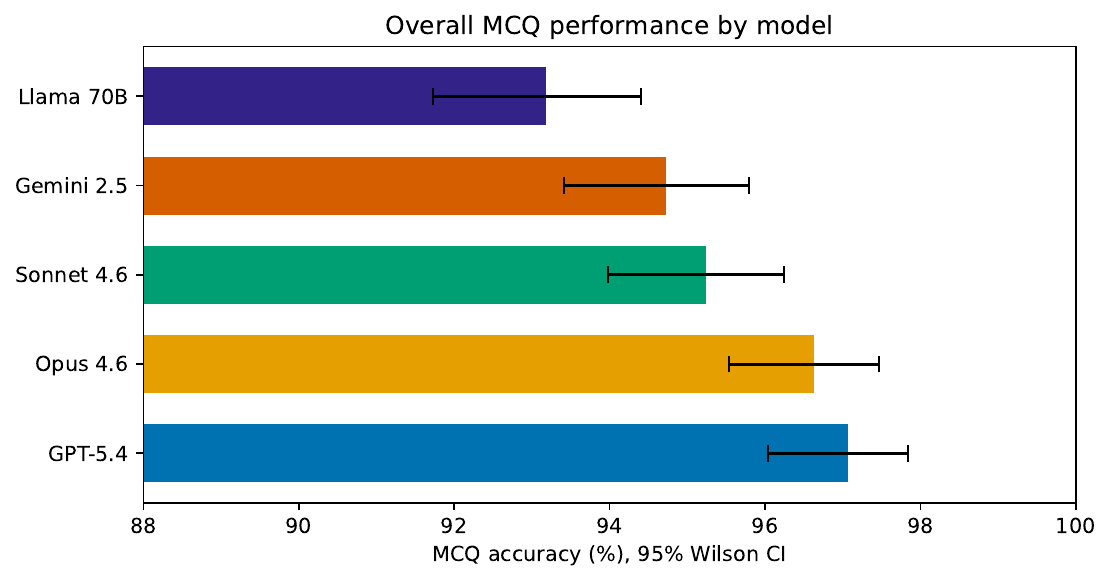}
\caption{Overall MCQ accuracy (\%) for five models with 95\% Wilson score confidence intervals (n = 1365). Overlapping intervals for GPT-5.4 and Opus 4.6 (P = .48) and for Sonnet 4.6 and Gemini 2.5 Pro (P = .44).}
\label{fig:fig_overall_accuracy_ci}
\end{figure}

\Needspace{8\baselineskip}
\textbf{Table 1.} Overall MCQ accuracy, correct/incorrect counts, and 95\% Wilson confidence intervals (n = 1365).

{\def\LTcaptype{none} 
\begin{longtable}[]{@{}lllll@{}}
\toprule\noalign{}
Model & Accuracy & Correct & Incorrect & 95\% CI \\
\midrule\noalign{}
\endhead
\bottomrule\noalign{}
\endlastfoot
GPT-5.4 & 97.1\% & 1325 & 40 & 96.0--97.8\% \\
Opus 4.6 & 96.6\% & 1319 & 46 & 95.5--97.5\% \\
Sonnet 4.6 & 95.2\% & 1300 & 65 & 94.0--96.2\% \\
Gemini 2.5 Pro & 94.7\% & 1293 & 72 & 93.4--95.8\% \\
Llama 3.3 70B & 93.2\% & 1272 & 93 & 91.7--94.4\% \\
\end{longtable}
}

The total spread across five model families was 3.9 percentage points (pp); among the four frontier models alone, 2.4 pp.~The top performers GPT-5.4 and Opus 4.6 were statistically indistinguishable (P = .48), as were Sonnet 4.6 and Gemini 2.5 Pro (P = .44), consistent with the overlapping Wilson CIs in Table 1. GPT-5.4 significantly outperformed both Sonnet 4.6 (P = .0018) and Gemini 2.5 Pro (P \textless{} .001). All four frontier models scored higher than Llama 3.3 70B (P \ensuremath{\leq} .030 for all pairs).

Tier-level results are presented in Table 2. Accuracy was stable between Tier 1 (n = 1134) and Tier 2 (n = 213) for most models. Gemini 2.5 Pro was the sole model whose Tier 2 accuracy exceeded Tier 1 (96.2\% vs 94.6\%); Gemini 2.5 Pro's mandatory chain-of-thought reasoning represents a methodological asymmetry relative to the other four models (see Methods). Tier 3 accuracy (n = 18) was lower for most models; results at this tier should be interpreted with caution given the small sample.

\textbf{Table 2.} MCQ accuracy (\%) by difficulty tier.

{\def\LTcaptype{none} 
\begin{longtable}[]{@{}
  >{\raggedright\arraybackslash}p{(\linewidth - 6\tabcolsep) * \real{0.2188}}
  >{\raggedright\arraybackslash}p{(\linewidth - 6\tabcolsep) * \real{0.2656}}
  >{\raggedright\arraybackslash}p{(\linewidth - 6\tabcolsep) * \real{0.2500}}
  >{\raggedright\arraybackslash}p{(\linewidth - 6\tabcolsep) * \real{0.2656}}@{}}
\toprule\noalign{}
\begin{minipage}[b]{\linewidth}\raggedright
Model
\end{minipage} & \begin{minipage}[b]{\linewidth}\raggedright
Tier 1 (n = 1134)
\end{minipage} & \begin{minipage}[b]{\linewidth}\raggedright
Tier 2 (n = 213)
\end{minipage} & \begin{minipage}[b]{\linewidth}\raggedright
Tier 3 (n = 18)*
\end{minipage} \\
\midrule\noalign{}
\endhead
\bottomrule\noalign{}
\endlastfoot
GPT-5.4 & 97.2 & 97.2 & 88.9 \\
Opus 4.6 & 96.6 & 96.2 & 100.0 \\
Sonnet 4.6 & 95.1 & 96.7 & 88.9 \\
Gemini 2.5 Pro & 94.6 & 96.2 & 83.3 \\
Llama 3.3 70B & 93.5 & 93.0 & 77.8 \\
\end{longtable}
}

*n = 18; interpret with caution.

Table 3 shows the main category-level pattern: GE scanner operations was the lowest-scoring category for every model (88.2--94.6\%) and showed the largest within-category spread (6.4 pp), whereas T1/T2 relaxation/contrast and parallel imaging were near ceiling. After Bonferroni correction, six pairwise comparisons reached P \textless{} .05. Three were in GE scanner operations, where GPT-5.4 outperformed Sonnet 4.6 (P = .032), Gemini 2.5 Pro (P = .001), and Llama 3.3 70B (P = .006). The remaining three, all involving Llama 3.3 70B, are reported in Appendix A. All other category-level pairwise comparisons were non-significant after correction.

\textbf{Table 3.} MCQ accuracy (\%) by category. \textbf{Bold} indicates lowest-scoring category per model.

{\def\LTcaptype{none} 
\begin{longtable}[]{@{}
  >{\raggedright\arraybackslash}p{(\linewidth - 10\tabcolsep) * \real{0.3537}}
  >{\raggedright\arraybackslash}p{(\linewidth - 10\tabcolsep) * \real{0.1220}}
  >{\raggedright\arraybackslash}p{(\linewidth - 10\tabcolsep) * \real{0.0976}}
  >{\raggedright\arraybackslash}p{(\linewidth - 10\tabcolsep) * \real{0.0976}}
  >{\raggedright\arraybackslash}p{(\linewidth - 10\tabcolsep) * \real{0.1707}}
  >{\raggedright\arraybackslash}p{(\linewidth - 10\tabcolsep) * \real{0.1585}}@{}}
\toprule\noalign{}
\begin{minipage}[b]{\linewidth}\raggedright
Category
\end{minipage} & \begin{minipage}[b]{\linewidth}\raggedright
Sonnet 4.6
\end{minipage} & \begin{minipage}[b]{\linewidth}\raggedright
Opus 4.6
\end{minipage} & \begin{minipage}[b]{\linewidth}\raggedright
GPT-5.4
\end{minipage} & \begin{minipage}[b]{\linewidth}\raggedright
Gemini 2.5 Pro
\end{minipage} & \begin{minipage}[b]{\linewidth}\raggedright
Llama 3.3 70B
\end{minipage} \\
\midrule\noalign{}
\endhead
\bottomrule\noalign{}
\endlastfoot
Artifacts & 97.5 & 98.8 & 100.0 & 98.8 & 97.5 \\
GE domain knowledge & 96.8 & 97.4 & 95.5 & 96.1 & 92.9 \\
GE scanner operations & \textbf{90.1} & \textbf{92.6} & \textbf{94.6} & \textbf{88.2} & \textbf{88.7} \\
k-space and image formation & 96.5 & 99.1 & 99.1 & 98.2 & 91.2 \\
Parallel imaging & 100.0 & 100.0 & 100.0 & 100.0 & 95.0 \\
Pulse sequences & 97.9 & 97.3 & 98.2 & 97.0 & 94.5 \\
Safety & 97.9 & 99.3 & 98.6 & 97.2 & 98.6 \\
SNR and image quality & 96.2 & 100.0 & 96.2 & 98.1 & 96.2 \\
T1/T2 relaxation and contrast & 97.0 & 100.0 & 100.0 & 100.0 & 98.5 \\
\end{longtable}
}

Four models were evaluated for run-to-run repeatability (Table 4). Perfect agreement is not expected because frontier LLMs exhibit small run-to-run variability even at temperature = 0, reflecting non-deterministic elements in GPU floating-point computation. All showed near-perfect agreement (kappa 0.983--0.998). Disagreements concentrated in GE scanner operations across models: Sonnet 4.6 (4 of 8 total disagreements), Llama 3.3 70B (4/7), and GPT-5.4 (8/17). Opus 4.6 was the most stable model, with only 2 total disagreements. The pattern is established across four model families. Gemini 2.5 Pro was not evaluated for repeatability because the API's daily quota limit (1000 requests/day) made a second full benchmark run impractical.

\textbf{Table 4.} Run-to-run repeatability (two runs per model, n = 1365).

{\def\LTcaptype{none} 
\begin{longtable}[]{@{}
  >{\raggedright\arraybackslash}p{(\linewidth - 8\tabcolsep) * \real{0.1757}}
  >{\raggedright\arraybackslash}p{(\linewidth - 8\tabcolsep) * \real{0.1216}}
  >{\raggedright\arraybackslash}p{(\linewidth - 8\tabcolsep) * \real{0.1757}}
  >{\raggedright\arraybackslash}p{(\linewidth - 8\tabcolsep) * \real{0.2568}}
  >{\raggedright\arraybackslash}p{(\linewidth - 8\tabcolsep) * \real{0.2703}}@{}}
\toprule\noalign{}
\begin{minipage}[b]{\linewidth}\raggedright
Model
\end{minipage} & \begin{minipage}[b]{\linewidth}\raggedright
Agreement
\end{minipage} & \begin{minipage}[b]{\linewidth}\raggedright
Cohen's kappa \cite{cohen_coefficient_1960}
\end{minipage} & \begin{minipage}[b]{\linewidth}\raggedright
Accuracy difference
\end{minipage} & \begin{minipage}[b]{\linewidth}\raggedright
GE ops disagreements
\end{minipage} \\
\midrule\noalign{}
\endhead
\bottomrule\noalign{}
\endlastfoot
Opus 4.6 & 99.85\% & 0.998 & 0.15\% & 1/2 \\
Llama 3.3 70B & 99.49\% & 0.993 & 0.07\% & 4/7 \\
Sonnet 4.6 & 99.41\% & 0.992 & 0.07\% & 4/8 \\
GPT-5.4 & 98.75\% & 0.983 & 0.15\% & 8/17 \\
\end{longtable}
}

Because the MCQ results compressed near ceiling, we next evaluated the same benchmark without answer options to test whether the narrow MCQ spread reflected recall or option recognition. A subset of four models was evaluated in a stem-only condition, in which MCQ options were removed and models generated free-text responses scored by an independent LLM judge (Grok). (Gemini 2.5 Pro excluded, see Appendix E.) Stem-only was analyzed as a complementary diagnostic format alongside the primary MCQ benchmark. Table 5 presents overall accuracy.

\textbf{Table 5.} MCQ vs stem-only accuracy.

{\def\LTcaptype{none} 
\begin{longtable}[]{@{}llll@{}}
\toprule\noalign{}
Model & MCQ & Stem-only & Difference \\
\midrule\noalign{}
\endhead
\bottomrule\noalign{}
\endlastfoot
Opus 4.6 & 96.6\% & 61.1\% & -35.5 pp \\
GPT-5.4 & 97.1\% & 60.3\% & -36.8 pp \\
Sonnet 4.6 & 95.2\% & 58.4\% & -36.8 pp \\
Llama 3.3 70B & 93.2\% & 37.1\% & -56.1 pp \\
\end{longtable}
}

Frontier models clustered tightly in stem-only (58.4--61.1\%, 2.7 pp spread), paralleling the narrow MCQ spread. Llama 3.3 70B showed a substantially larger difference (56.1 pp vs 35.5--36.8 pp for frontier models), falling to 37.1\% (Figure 3). At the category level, stem-only accuracy ranged from near-ceiling on T1/T2 relaxation and contrast (92.5--95.5\% for frontier models) to near-floor on GE scanner operations (13.8--29.8\% across all models), indicating substantially weaker free-text recall for this vendor-specific content than MCQ performance would suggest (Figure 2). The full category-level stem-only results are shown in Figure 4 with numeric values in Table B1 (Appendix B). Human spot-check of Grok judgments (n = 136 Sonnet 4.6 responses) yielded 89.7\% agreement, with disagreements predominantly reflecting judge strictness. A post-hoc linguistic uncertainty analysis examined whether hedging language in free-text responses predicted incorrectness across all four stem-only models (n = 5460 responses total; see Appendix I). Stem-only results are interpreted as descriptive complementary evidence.

\begin{figure}[htbp]
\centering
\includegraphics[width=\linewidth]{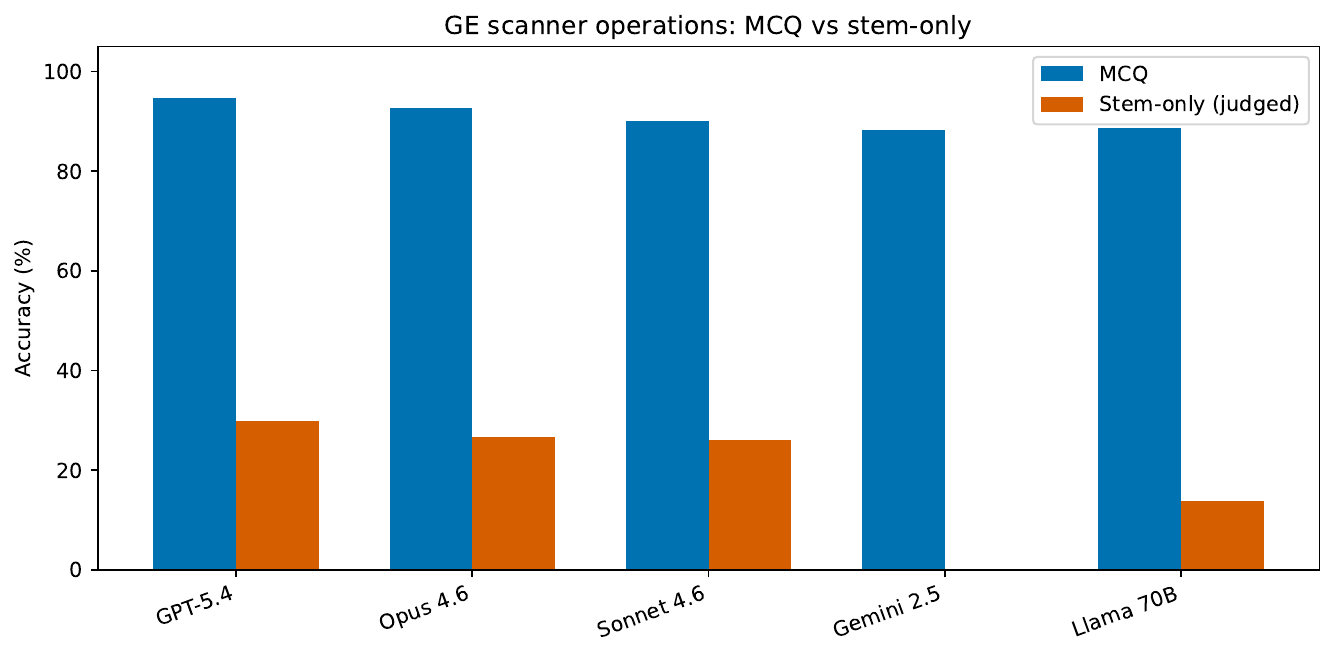}
\caption{MCQ versus stem-only accuracy for the GE scanner operations category. MCQ accuracy ranged from 88.2\% to 94.6\% across five models, and stem-only accuracy collapsed to 13.8--29.8\% for the four evaluated models. Gemini 2.5 Pro was excluded from stem-only evaluation due to mandatory reasoning tokens.}
\label{fig:fig_ge_ops_mcq_vs_stem}
\end{figure}

\begin{figure}[htbp]
\centering
\includegraphics[width=\linewidth]{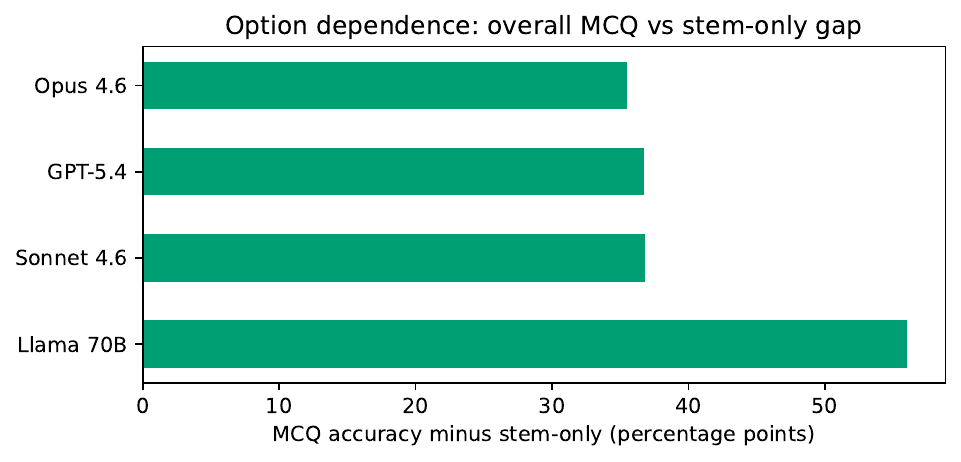}
\caption{Overall MCQ-minus-stem-only accuracy difference (pp) for four models. Frontier models clustered at 35.5--36.8 pp, and Llama 3.3 70B showed a substantially larger difference (56.1 pp), indicating greater option-dependence. Gemini 2.5 Pro was excluded from stem-only evaluation.}
\label{fig:fig_mcq_minus_stem_delta}
\end{figure}

\begin{figure}[htbp]
\centering
\includegraphics[width=\linewidth]{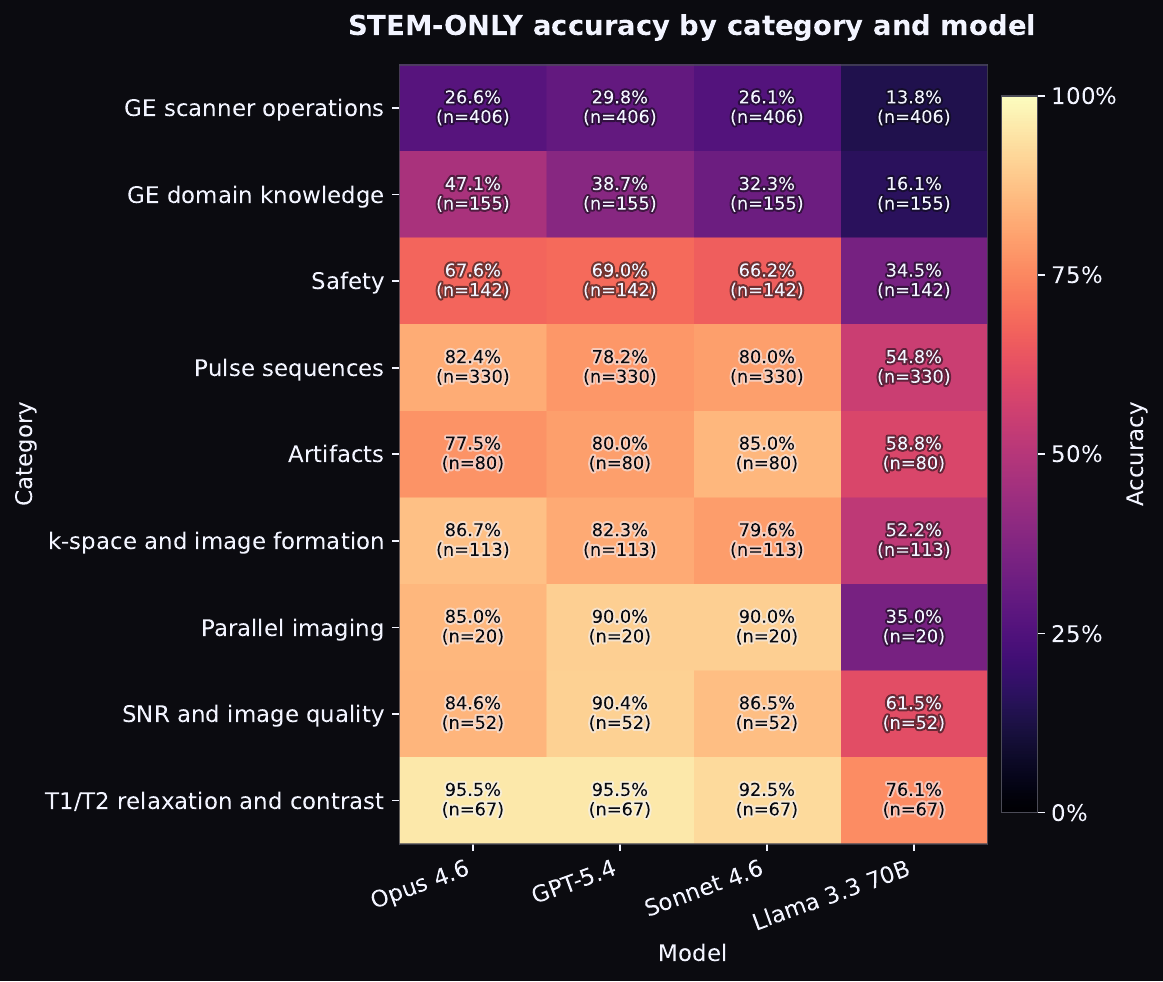}
\caption{Stem-only accuracy (\%) by category and model (heatmap; n per cell shown). Categories are ordered from lowest to highest stem-only accuracy (top to bottom), sorted by the GPT-5.4 column. The gradient from T1/T2 relaxation and contrast (92.5--95.5\% for frontier models) to GE scanner operations (13.8--29.8\%) illustrated the knowledge gradient from physics fundamentals to vendor-specific operational content. Gemini 2.5 Pro was excluded from stem-only evaluation.}
\label{fig:stem_only_category_model_heatmap}
\end{figure}

\FloatBarrier

The secondary primed diagnostic condition presented each model with an incorrect answer and asked whether it agreed (Table 6). Primed accuracy exceeded unprimed for every model overall, indicating that answer-cueing dominated this condition. Llama 3.3 70B showed a qualitatively different tier-level pattern: accuracy declined under priming at Tier 2 (-3.3 pp) and Tier 3 (-5.6 pp; n = 18 caveat), while frontier models improved at all tiers. Category-level primed results are in Appendix C.

\textbf{Table 6.} Unprimed vs primed stem-only accuracy.

\Needspace{12\baselineskip}
{\def\LTcaptype{none} 
\begin{longtable}[]{@{}llll@{}}
\toprule\noalign{}
Model & Unprimed & Primed & Difference \\
\midrule\noalign{}
\endhead
\bottomrule\noalign{}
\endlastfoot
GPT-5.4 & 60.3\% & 67.4\% & +7.1 pp \\
Opus 4.6 & 61.1\% & 65.9\% & +4.8 pp \\
Sonnet 4.6 & 58.4\% & 62.9\% & +4.5 pp \\
Llama 3.3 70B & 37.1\% & 39.6\% & +2.4 pp \\
\end{longtable}
}

Misconception item results (n = 10) are reported in Appendix D.

\subsection{Discussion}\label{discussion}

MRI-Eval changes how high MRI LLM scores should be interpreted: in this benchmark, near-ceiling MCQ performance did not translate into strong free-text recall for GE-specific scanner operational content. GE scanner operations was already the hardest MCQ category for every model, but the stem-only results make the practical boundary clearer by showing how sharply performance falls once answer options are removed (Figure 2). On that category, GPT-5.4 significantly outperformed Sonnet 4.6, Gemini 2.5 Pro, and Llama 3.3 70B after multiple-comparison correction, but not Opus 4.6, which suggests somewhat deeper GE-specific knowledge in that dimension than the narrow frontier MCQ margins imply. Even so, no model approached strong operational recall in free text. A representative qualitative example illustrates the issue: on a stem-only question about GE parallel imaging, Sonnet 4.6 produced GRAPPA instead of ARC, retrieving a salient cross-vendor association rather than the GE-specific term required by the benchmark. GE scanner operations and GE domain knowledge together constitute 41\% of the benchmark (561/1365 items), reflecting a GE research environment (GE SIGNA Premier 3T installation). This boundary in performance supports caution against relying on raw LLM outputs for vendor-specific protocol guidance in research MRI contexts and motivates evaluation of retrieval-augmented or document-grounded approaches, which remain untested here.

The narrow MCQ spread among frontier models does not imply equivalent capability. GPT-5.4 and Opus 4.6 were statistically indistinguishable overall, as were Sonnet 4.6 and Gemini 2.5 Pro, which is consistent with two distinct capability tiers collapsing to similar MCQ scores once core domain knowledge is largely internalized. That pattern fits a ceiling-compression effect: MCQ loses discriminative power near the top (Figure 1), and harder formats are needed to separate models. Gemini 2.5 Pro alone showed Tier 2 accuracy above Tier 1, though mandatory internal chain-of-thought for Gemini 2.5 Pro is a methodological asymmetry that prevents attributing that inversion cleanly to a reasoning advantage (Appendix E).

The large MCQ-versus-stem-only gap for frontier models (Figure 3) implies that a substantial share of correct MCQ answers reflect option recognition rather than recall. The similarity of the drop across three frontier families points to a structural feature of MCQ rather than idiosyncratic model behavior, although the pervasive option-length imbalance documented in Methods means the MCQ arm likely contains both answer-recognition and option-length cues. Stem-only performance remained tightly clustered among frontier models, paralleling narrow MCQ spread and showing that compression persists without options. Category-level stem-only performance ranged from strong on physics fundamentals to very weak on GE scanner operations (Figure 4; Table B1). An analysis found no reliable relationship between answer hedging language (e.g.~``maybe'', ``possibly'') and actual incorrectness, confirming that answer uncertainty cannot serve as a signal for identifying unreliable answers. Together, these findings support including stem-only or other open-ended formats alongside MCQ in future domain-specific benchmarks, ideally with stronger validation and inferential treatment than was feasible here.

Llama 3.3 70B differed qualitatively: its MCQ-to-stem-only gap was much larger than the frontier model group, indicating greater option-dependence. Overall stem-only accuracy was far lower than for frontier models; safety exhibited the largest category-level MCQ-versus-stem-only drop for Llama 3.3 70B (Appendix B). Even where Llama 3.3 70B's MCQ scores resembled frontier models', underlying knowledge was weaker, so MCQ overstates its competency more than it does for frontier systems.

Under the secondary primed diagnostic condition, every model scored higher when primed than when unprimed (Table 6). That overall positive change indicates that answer-cueing dominated the condition for frontier models. The most interpretable finding in the data is Llama 3.3 70B alone showing reduced accuracy under priming at Tier 2 and Tier 3 (while frontier models improved at every tier), consistent with weaker models partially accepting incorrect primes when question difficulty exceeds knowledge. Category-level primed contrasts are summarized in Appendix C. Consequently we must be mindful whether the model is primed or cued by questions with incorrect information.

Several limitations constrain interpretation. Tier 3 comprised only 18 items, insufficient for reliable statistical inference, hence those results are indicative only. No human testing was evaluated, so the reported accuracies should be interpreted as relative model performance within this benchmark rather than as direct evidence of expert-level or practice-ready competency.\footnote{``It is one thing to remember, another to know. Remembering is merely safeguarding something entrusted to the memory; knowing, however, means making everything your own.'' --- Seneca, Letter 33.} A post-hoc option-length audit found that 84.9\% of MCQ items had a correct option exceeding 1.3 times the mean distractor length, with a consistent positive accuracy gap between length-flagged and length-balanced subsets across all five models (+8.3 to +20.0 pp). Given the prevalence of this imbalance, option-length cues should be regarded as an interpretive limitation of the MCQ arm rather than a minor subset issue. While the observational design cannot establish causality, the prevalence and direction of the association make it difficult to treat MCQ accuracy as a clean measure of knowledge alone. The stem-only evaluation, which removes all options, provides only a partial control because it eliminates answer options entirely rather than isolating option-length effects specifically. The consistent cross-model ranking and tight frontier clustering in stem-only results argue that the principal conclusions are not solely artifacts of MCQ cueing, but they do not remove the MCQ confound. The Grok judge was validated on a single-model subsample (n = 136 Sonnet 4.6 responses) with 89.7\% agreement. Cross-model and category-level stem-only comparisons rest on the assumption that judge bias is uniform across models, which was not independently tested. Systematic conservative bias means absolute stem-only accuracy figures are modestly underestimated, though relative model comparisons remain valid under the uniform-bias assumption (Appendix H). Bonferroni correction was applied to MCQ category-level pairwise tests, but stem-only pairwise contrasts were not formally tested. Primed category differences, tier-level observations, option-length analyses, and linguistic-uncertainty results were also not subject to formal multiple-comparison correction across those contrasts. The benchmark was authored and validated by a single domain expert. While AI-assisted screening provided a partial external check, independent multi-rater validation would strengthen confidence in item quality. Results reflect parametric (internal model) knowledge only, and potential enhancements with internet or document search were not tested.

Planned extensions center on two lines of work. The first is evaluation of retrieval-augmented systems using vendor documentation as the retrieval sources, directly motivated by the GE scanner operations stem-only floor. The second is purpose-built sycophancy evaluation with confidence rating elicitation, replacing the diagnostic primed template with a design intended to separate agreement-with-user from answer-cueing effects.

\subsection{Conclusion}\label{conclusion}

MRI-Eval suggests that the main value of high-scoring MRI MCQ benchmarks is comparative rather than for establishing true domain competence. Across models, standard MCQ evaluation compressed performance into a narrow range, while stem-only evaluation exposed a much sharper boundary between broad MRI knowledge and GE-specific operational recall. That contrast was most consequential in GE scanner operations, where removing answer options revealed that none of the tested models could reliably generate strong free-text guidance.

The benchmark therefore supports two practical conclusions. First, strong MCQ performance should not be treated as evidence of deployment-ready vendor-specific knowledge. Second, future progress in this domain is unlikely to be captured by MCQ benchmarking alone. Harder open-ended evaluation is a better next target for assessment. These findings also motivate testing retrieval-augmented or document-grounded systems as a next step. The primed diagnostic condition further suggests that incorrect-answer prompts mainly functioned as answer cues rather than as a clean stress test of agreement behavior, with the clearest vulnerability appearing in the weaker open-weight model at higher difficulty. Taken together, MRI-Eval benchmarks relative model performance and format sensitivity within a GE MRI setting, but it does not provide a human-reference calibration for absolute competency.

\renewcommand{\refname}{References}
\bibliographystyle{plain}
\bibliography{bibliography}

\subsection{Appendix A. Complete Pairwise Statistical Comparisons}\label{appendix-a.-complete-pairwise-statistical-comparisons}

The main text reports three of the six category-level pairwise comparisons that reached P \textless{} .05 after Bonferroni correction; those three were all in the GE scanner operations category. The remaining three reached significance in other categories:

\begin{itemize}
\tightlist
\item
  k-space and image formation: Opus 4.6 vs Llama 3.3 70B (P = .035); GPT-5.4 vs Llama 3.3 70B (P = .035)
\item
  Pulse sequences: GPT-5.4 vs Llama 3.3 70B (P = .038)
\end{itemize}

All three involved Llama 3.3 70B as the lower-performing model, confirming that Llama 3.3 70B's deficit relative to frontier models extended beyond GE-specific content into core physics categories. All Bonferroni-corrected P values were derived from the p\_value\_bonferroni column of \texttt{results/stats/mcnemar\_pairwise.csv}. The full pairwise comparison table, including raw and adjusted P values for all model pairs across all nine categories, is available in the released repository artifacts.

\subsection{Appendix B. Stem-Only Category-Level Results}\label{appendix-b.-stem-only-category-level-results}

Table B1 presents stem-only accuracy by category for the four evaluated models (Gemini 2.5 Pro excluded; see Methods).

\textbf{Table B1.} Stem-only accuracy (\%) by category (n = 1365 items per model; scored by Grok judge).

{\def\LTcaptype{none} 
\begin{longtable}[]{@{}
  >{\raggedright\arraybackslash}p{(\linewidth - 8\tabcolsep) * \real{0.4328}}
  >{\raggedright\arraybackslash}p{(\linewidth - 8\tabcolsep) * \real{0.1194}}
  >{\raggedright\arraybackslash}p{(\linewidth - 8\tabcolsep) * \real{0.1045}}
  >{\raggedright\arraybackslash}p{(\linewidth - 8\tabcolsep) * \real{0.1493}}
  >{\raggedright\arraybackslash}p{(\linewidth - 8\tabcolsep) * \real{0.1940}}@{}}
\toprule\noalign{}
\begin{minipage}[b]{\linewidth}\raggedright
Category
\end{minipage} & \begin{minipage}[b]{\linewidth}\raggedright
Opus 4.6
\end{minipage} & \begin{minipage}[b]{\linewidth}\raggedright
GPT-5.4
\end{minipage} & \begin{minipage}[b]{\linewidth}\raggedright
Sonnet 4.6
\end{minipage} & \begin{minipage}[b]{\linewidth}\raggedright
Llama 3.3 70B
\end{minipage} \\
\midrule\noalign{}
\endhead
\bottomrule\noalign{}
\endlastfoot
T1/T2 relaxation and contrast & 95.5 & 95.5 & 92.5 & 76.1 \\
Parallel imaging & 85.0 & 90.0 & 90.0 & 35.0 \\
SNR and image quality & 84.6 & 90.4 & 86.5 & 61.5 \\
k-space and image formation & 86.7 & 82.3 & 79.6 & 52.2 \\
Pulse sequences & 82.4 & 78.2 & 80.0 & 54.8 \\
Artifacts & 77.5 & 80.0 & 85.0 & 58.8 \\
Safety & 67.6 & 69.0 & 66.2 & 34.5 \\
GE domain knowledge & 47.1 & 38.7 & 32.3 & 16.1 \\
GE scanner operations & 26.6 & 29.8 & 26.1 & 13.8 \\
\end{longtable}
}

Safety produced the largest individual category difference for any model: Llama 3.3 70B dropped from 98.6\% MCQ accuracy to 34.5\% stem-only (64 pp), indicating that its safety knowledge was highly option-dependent despite near-ceiling MCQ performance. GE scanner operations remained the floor across all four models (13.8--29.8\%). No model demonstrated strong free-text recall of GE-specific operational content without MCQ options.

\subsubsection{Qualitative Examples}\label{qualitative-examples}

Two additional illustrative cases from Sonnet 4.6 stem-only responses are presented below.

\textbf{ID 0055 (Terminology).} The model used a fictional acronym (EPO) instead of the correct term (EMO) but provided a conceptually correct answer. The Grok judge scored this as correct, consistent with the rubric distinguishing terminology errors from knowledge errors.

\textbf{ID 0471 (Artifacts).} The model produced a confident but incorrect three-artifact classification in free text whereas the GE manual specifies a different taxonomy. This type of confidently wrong answer was undetectable in the MCQ condition, where the correct option was available for selection.

\subsection{Appendix C. Primed Evaluation Category and Tier Details}\label{appendix-c.-primed-evaluation-category-and-tier-details}

Table C1 supplements Table 6 in the main text with selected category-level primed differences representing the largest positive and negative shifts observed.

\textbf{Table C1.} Selected category-level primed vs unprimed stem-only accuracy differences.

{\def\LTcaptype{none} 
\begin{longtable}[]{@{}lllll@{}}
\toprule\noalign{}
Model & Category & Unprimed & Primed & difference \\
\midrule\noalign{}
\endhead
\bottomrule\noalign{}
\endlastfoot
Llama 3.3 & Parallel imaging & 35.0\% & 60.0\% & +25.0 pp \\
Sonnet 4.6 & GE domain knowledge & 32.3\% & 46.5\% & +14.2 pp \\
Llama 3.3 & Artifacts & 58.8\% & 47.5\% & -11.3 pp \\
Sonnet 4.6 & Artifacts & 85.0\% & 75.0\% & -10.0 pp \\
\end{longtable}
}

Llama 3.3 70B was the only model whose accuracy declined under priming at higher difficulty tiers: Tier 2 decreased by 3.3 pp and Tier 3 by 5.6 pp (n = 18; interpret with caution), while Tier 1 improved slightly. All frontier models improved at every tier under the primed condition. Llama 3.3 70B's tier-level pattern is consistent with partial acceptance of incorrect primes on harder questions (the sycophancy-oriented behavior the secondary diagnostic condition was motivated to probe) and contrasts with the error-recognition scaffolding effect that predominated for frontier models, as observed under the Basic Nudge condition \cite{christophe_overalignment_2026}.

\subsection{Appendix D. Misconception Question Results}\label{appendix-d.-misconception-question-results}

Table D1 presents the full misconception results for all five models (n = 10 items; counts only, insufficient for statistical testing).

\textbf{Table D1.} Misconception question accuracy by model (n = 10; counts only, insufficient for statistical testing).

{\def\LTcaptype{none} 
\begin{longtable}[]{@{}llllll@{}}
\toprule\noalign{}
Model & n & Correct & Incorrect & Accuracy & Wrong IDs \\
\midrule\noalign{}
\endhead
\bottomrule\noalign{}
\endlastfoot
Opus 4.6 & 10 & 10 & 0 & 100\% & - \\
Sonnet 4.6 & 10 & 9 & 1 & 90\% & 1441 \\
GPT-5.4 & 10 & 9 & 1 & 90\% & 1441 \\
Gemini 2.5 Pro & 10 & 9 & 1 & 90\% & 1441 \\
Llama 3.3 70B & 10 & 8 & 2 & 80\% & 1431, 1441 \\
\end{longtable}
}

Opus 4.6 was the only model that correctly handled all 10 items. ID 1441 was failed by every model except Opus 4.6; ID 1431 was an additional failure for Llama 3.3 70B only.

\subsection{Appendix E. Gemini 2.5 Pro: Configuration Details and Tier 2 Inversion}\label{appendix-e.-gemini-2.5-pro-configuration-details-and-tier-2-inversion}

\subsubsection{E.1 Configuration and Replication Notes}\label{e.1-configuration-and-replication-notes}

Gemini 2.5 Pro was configured with thinking\_budget = 512 (a soft cap on internal reasoning tokens) and max\_output\_tokens = 1024. Without the thinking\_budget constraint, reasoning tokens exhausted the shared output budget in initial runs, producing 52/1365 (3.8\%) empty responses. Those results were discarded and the corrected configuration was rerun before any analysis entered the manuscript. With the corrected configuration, Gemini 2.5 Pro averaged 398 output tokens per question compared with approximately 5 tokens per question for the four other models. This token budget partitioning behavior-in which reasoning tokens compete with response tokens within a shared output allocation-is not prominently documented by Google and is noted as a replication hazard for researchers using the Gemini API for structured evaluations.

\subsubsection{E.2 Tier 2 Accuracy Inversion}\label{e.2-tier-2-accuracy-inversion}

Gemini 2.5 Pro was the only model whose Tier 2 accuracy exceeded Tier 1 (96.2\% vs 94.6\%). This pattern is consistent with mandatory internal chain-of-thought reasoning providing greater benefit on cross-concept Tier 2 items, which require synthesis across multiple concepts, than on single-concept Tier 1 recall items. However, the methodological asymmetry-Gemini 2.5 Pro operated in implicit chain-of-thought mode while all other models responded in direct letter-only mode-means this comparison is not fully controlled. The inversion may reflect a genuine reasoning advantage on integrative items, a confound introduced by the evaluation asymmetry, or both. This finding motivates explicit chain-of-thought prompting experiments for other models on Tier 2 items in future controlled comparisons across model families. Planned extensions emphasize RAG and purpose-built sycophancy evaluation (see Discussion); Tier-2 CoT comparisons are complementary follow-on work.

\subsection{Appendix F. Benchmark Construction and Screening Details}\label{appendix-f.-benchmark-construction-and-screening-details}

All 1365 scored benchmark items (1355 MCQ + 10 misconception) were confirmed stem-only compatible under the harness loader. Six MCQ items and all 10 misconception items used stem overrides for reframing in the free-text evaluation conditions. No items were excluded from stem-only evaluation for incompatibility.

The YAML question bank underwent nine structured AI-assisted screening passes followed by author review against the primary sources. Across those passes, flagged items were triaged and resolved, with a small number retired before evaluation because they were duplicates, near-duplicates, out of scope for the benchmark, or required external clinical verification beyond the source set available at authoring time. One ophthalmic-safety item was explicitly excluded from reported runs pending confirmation against current American College of Radiology (ACR) guidance.

Additional revisions included category retagging, stem and distractor cleanup, and explanation edits. The repository audit trail retains those low-level screening notes, while the manuscript reports only the reader-facing outcome: a final scored benchmark of 1365 items. The repository also contains short-answer and scenario items that were developed during benchmark construction but were not part of the evaluated MCQ-plus-misconception set.

\subsection{Appendix G. Evaluation Harness: Reporting Runs and Statistical Implementation}\label{appendix-g.-evaluation-harness-reporting-runs-and-statistical-implementation}

\subsubsection{G.1 Reporting Artifacts}\label{g.1-reporting-artifacts}

MCQ results reported in Tables 1-3 were generated from one designated reporting run per model. Repeatability reruns and sensitivity runs were kept separate from those reporting runs. Exact run identifiers, commands, and exported CSV artifacts are provided in the released repository.

\subsubsection{G.2 Statistical Implementation Notes}\label{g.2-statistical-implementation-notes}

Confidence intervals for all reported proportions (overall, per tier, per category) were two-sided Wilson score 95\% intervals, computed using \texttt{statsmodels} with \texttt{method="wilson"} and \texttt{alpha=0.05}. No normal approximation was used in the pipeline.

McNemar's test \cite{mcnemar_note_1947} for paired model comparisons was implemented via \texttt{statsmodels.stats.contingency\_tables.mcnemar}. The exact test was applied when the number of discordant pairs was fewer than 25 (no continuity correction); the asymptotic test with continuity correction was applied when discordant pairs numbered 25 or more. For each unordered model pair, nine category-level McNemar tests were Bonferroni-adjusted within that pair (adjusted P = min(1, 9p)). Overall pairwise McNemar P values reported in \texttt{mcnemar\_pairwise.csv} were not subject to this Bonferroni multiplier; only category-level P values within each pair were adjusted.

\subsection{Appendix H. Judge Validation and Vendor Context Preamble}\label{appendix-h.-judge-validation-and-vendor-context-preamble}

\subsubsection{H.1 Judge Disagreement Analysis}\label{h.1-judge-disagreement-analysis}

Human validation of the Grok (xAI) stem-only judge was performed on a subset of Sonnet 4.6 responses (n = 136, comprising all Tier 2 and Tier 3 items plus a random Tier 1 sample). Of 14 judge-human disagreements, 13 reflected Grok strictness (judge scored incorrect where the human scored correct) and 1 reflected leniency (judge scored correct where the human scored incorrect). Reported stem-only accuracy figures are therefore modestly conservative, with true accuracy estimated at approximately 2--3 pp higher. Relative model comparisons remain valid because the systematic strictness applied uniformly across all models, but the limited single-model validation is one reason the main text treats stem-only results as complementary descriptive evidence rather than as a fully symmetric inferential endpoint.

\subsubsection{H.2 Vendor Context Preamble Effect}\label{h.2-vendor-context-preamble-effect}

For questions in the GE scanner operations and GE domain knowledge categories, a vendor-context preamble was appended to the stem-only prompt: ``Where this question refers to a specific MRI system, scanner platform, or software environment, assume it is a GE SIGNA Premier 3T MRI system.'' This preamble reduced context-hedging responses (in which the model declined to answer due to unspecified vendor context) from 1.7\% to 0.07\% of items in those categories.

\subsection{Appendix I. Linguistic Uncertainty Analysis}\label{appendix-i.-linguistic-uncertainty-analysis}

A linguistic uncertainty analysis across all four stem-only models (n = 5460 responses total) examined whether hedging language in free-text responses predicted incorrectness. The analysis found no reliable relationship (AUC \textasciitilde0.50 across all models), indicating that model-generated free-text responses sounded equally confident regardless of whether the answer was correct or incorrect. Verbal uncertainty does not serve as a usable signal for identifying unreliable answers; users cannot distinguish correct from incorrect stem-only responses on the basis of expressed confidence or hedging alone. This is consistent with prior work showing that verbal uncertainty is a poor proxy for LLM calibration on factual questions \cite{xiong_can_2024}.

\subsection{Appendix J. Benchmark Cost}\label{appendix-j.-benchmark-cost}

Total API costs for all benchmark runs and associated development and pilot work were approximately \$62.10 USD; the Gemini total includes the discarded initial misconfiguration run and pilot tests.

\textbf{Table J1.} Per-model API costs.

{\def\LTcaptype{none} 
\begin{longtable}[]{@{}ll@{}}
\toprule\noalign{}
Model / Provider & Cost (USD) \\
\midrule\noalign{}
\endhead
\bottomrule\noalign{}
\endlastfoot
Gemini 2.5 Pro (Google AI Studio) & \$30.66 \\
Opus 4.6 (Anthropic) & \$13.81 \\
Sonnet 4.6 (Anthropic) & \$11.75 \\
GPT-5.4 (OpenAI) & \$4.88 \\
Grok (judge) (xAI) & \$0.98 \\
Llama 3.3 70B & \$0.62 \\
\end{longtable}
}

\end{document}